\documentclass[journal]{IEEEtran}
\usepackage{cite}
\usepackage{amsmath,amssymb,amsfonts}
\usepackage{graphicx}
\usepackage{textcomp}
\usepackage{xcolor}
\usepackage{hyperref}
\usepackage{amsmath}
\usepackage{autobreak}
\usepackage{amsfonts}
\usepackage{bm}
\usepackage{bbm}
\usepackage{algorithm}  
\usepackage{algorithmicx}  
\usepackage{algpseudocode}
\usepackage{stfloats}
\usepackage{cuted}
\usepackage{color}
\usepackage{xcolor}
\usepackage{subfigure}
\usepackage{cite}
\usepackage{booktabs} 
\usepackage{soul}
\usepackage{arydshln}
\usepackage{cases}
\definecolor{red}{RGB}{0,0,0}

\newcommand{\tr}{\text{tr}}

\newcommand{\red}[1]{{\color{red}#1}}

\ifCLASSINFOpdf
\makeatletter
\renewcommand{\maketag@@@}[1]{\hbox{\m@th\normalsize\normalfont#1}}%
\makeatother
\else
\fi
\begin{document}
		%
		\title{Bare Demo of IEEEtran.cls\\ for IEEE Journals}
		%
		%
		%
		
		\author{Wanting Lyu,~Songjie Yang,~Yue Xiu,~Xinyi Chen,~Zhongpei Zhang,~\IEEEmembership{Member,~IEEE},\\Chadi Assi,~\IEEEmembership{Fellow,~IEEE}, and Chau Yuen,~\IEEEmembership{Fellow,~IEEE} 
			
			\thanks{Wanting Lyu, Songjie Yang, Xinyi Chen and Zhongpei Zhang are with National Key Laboratory of Wireless Communications, University of Electronic Science and Technology of China, Chengdu 611731, China (E-mail: lyuwanting@yeah.net; yangsongjie@std.uestc.edu.cn; chenxinyi@std.uestc.edu.cn; zhangzp@uestc.edu.cn).
        
            Yue Xiu is with College of Air Traffic, Civil Aviation Flight University of China, Guanghan 618311, China (xiuyue12345678@163.com).
				
			Chadi Assi is with Concordia University, Montreal, Quebec, H3G 1M8, Canada (email:assi@ciise.concordia.ca).
				
			Chau Yuen is with the School of Electrical and Electronics Engineering, Nanyang Technological University, 639798 Singapore (E-mail: chau.yuen@ntu.edu.sg).}
		\vspace{-1cm}}

	\title{Dual-Robust Integrated Sensing and Communication: Beamforming under CSI Imperfection and Location Uncertainty }
	
	\maketitle
	
	\begin{abstract}
		A dual-robust design of beamforming is investigated in an integrated sensing and communication (ISAC) system. \red{
        Existing research on robust ISAC waveform design, while proposing solutions to imperfect channel state information (CSI), generally depends on prior knowledge of the target's approximate location to design waveforms. This approach, however, limits the precision in sensing the target's exact location.} In this paper, considering both CSI imperfection and target location uncertainty, a novel framework of joint robust optimization is proposed by maximizing the weighted sum of worst-case data rate and beampattern gain. To address this challenging problem, we propose an efficient two-layer iteration algorithm based on S-Procedure and convex hull. Finally, numerical results verify the effectiveness and performance improvement of our dual-robust algorithm, as well as the trade-off between communication and sensing performance. 
	\end{abstract}
	

	%
	\IEEEpeerreviewmaketitle

	\section{Introduction}

	\IEEEPARstart{I}{ntegrated} sensing and communication (ISAC) has emerged as a promising technology in future mobile networks. In numerous application scenarios, including internet-of-vehicles, smart homes, and environmental monitoring, the incorporation of environmental sensing into the communication stage has significantly evolved, enhancing overall system functionality and efficiency. \cite{9737357}. With an integrated hardware platform for both sensing and communication, ISAC systems offers mutual benefits to both functionalities, resulting in reduced hardware and energy costs \cite{10124135}. Due to these promising advantages, ISAC has been exploited in emerging technologies, such as reconfigurable intelligent surfaces (RIS) \cite{9724202}, non-orthogonal multiple access (NOMA) \cite{9668964}, and unmanned aerial vehicles (UAV) \cite{10098686}.
	
	In recent literature, there have been numerous researches dedicated to explore the joint beamforming design for ISAC system \cite{8288677,9729809,9868348,9729741,9852716,PYZ}. Authors in \cite{9729809} investigated a beampattern matching problem under the constraints guaranteeing SINR to be greater than the threshold. In contrast, in \cite{9729741} and \cite{9852716}, communication performance was the objective for RIS-aided beamforming optimization, guaranteeing the sensing SINR or beampattern matching error instead. To flexibly control the balance between sensing and communication priorities, a weighting factor was introduced to combine both metrics within the objective function \cite{PYZ}. In \cite{9668964}, the weighted sum of communication SINR and beampattern gain was maximized in a NOMA empowered ISAC system, while maintaining fairness constraints for all users and targets.
	
	\red{However, the above works of ISAC waveform design were all based on estimated CSI and the last known target locations, and assumed that the information is perfect at the BS. } \red{To address the  channel estimation errors, a few works have been dedicated into robust beamforming design for ISAC systems \cite{2023arXiv230307652B,2023arXiv231100071W}. In \cite{2023arXiv230307652B}, a robust beamforming design approach was proposed to tackle the imperfect CSI in ISAC, where sensing beampattern at the dedicated target location was maximized with outage probability constraint for communication. Also, channel estimation error is considered in \cite{2023arXiv231100071W}, through minimizing the worst-case multi-user interference energy. Additionally, robust beamforming design for ISAC has been exploited with other techniques, such as physical layer security \cite{10153696} and reconfigurable intelligent surface (RIS) \cite{10056405}. Despite the satisfactory results in the above works, most of them exclusively focused on communication channel estimation error, and ignored the target location uncertainty during waveform design. In addition, the objective function usually considered solely communication or sensing performance, limiting the flexibility to adjust priorities between the two functions.}
	
	In this paper, we propose a novel framework for dual-robust beamforming in ISAC. Our approach maximizes a combined metric of worst-case sum rate and beampattern gain, accounting for both CSI imperfection and target location uncertainty for beamforming optimization. To address the non-convex max-min problem, we propose an efficient robust algorithm based on the S-Procedure and convex hull. Simulation results demonstrate considerable performance improvement of 82\% for our dual-robust beamforming approach compared to non-robust designs considering CSI and location errors. We also investigate the trade-off between sensing and communication functions by adjusting the weighting factor.

	

	\section{System Model}
	
	\subsection{Signal Model}
	\begin{figure}[t]
		\centering
		\includegraphics[width=0.45\linewidth]{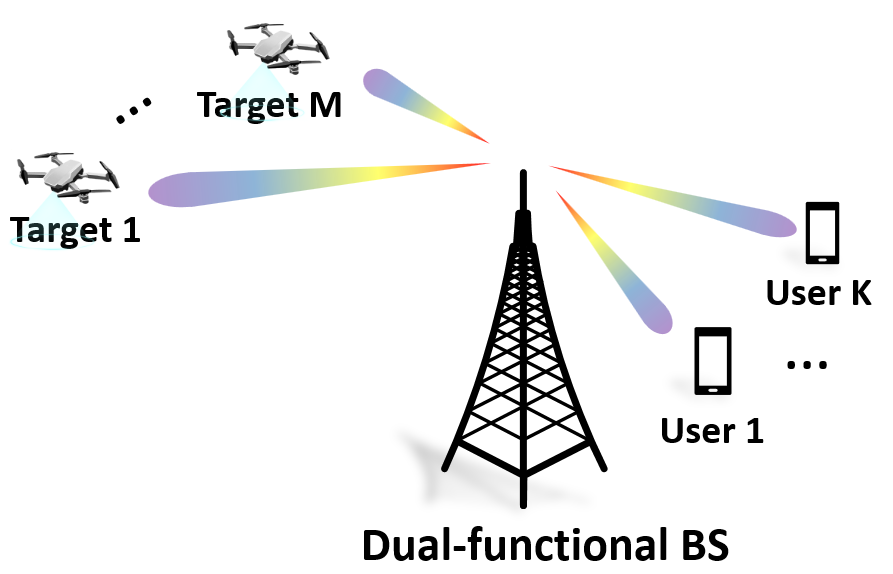}
		\caption{System model of the ISAC system}
		\label{system model}
	\end{figure}
	
	A DFRC BS is serving $K$ single-antenna communication users and sensing $M$ radar targets as shown in Fig. \ref{system model}. The BS is equipped with a fully-digital uniform linear array (ULA) with $N_t$ transmit antennas. Without loss of generality, the space between the adjacent elements for both the transmit array is $d_0 = \frac{\lambda}{2}$, where $\lambda$ denotes the wavelength. The users and the targets are assumed to be sufficiently far from the BS, and thus the target can be viewed as points \cite{8288677}.

	\red{We assume $K + M$ baseband data streams transmitted from the BS for communication and sensing, denoted as $\mathbf s \in \mathbb C^{(K+M)\times 1}$, where $K$ streams are dedicated for communication, and all signals can be utilized for sensing}. The transmitted signal $\mathbf s$ satisfies independent complex Gaussian distribution such that $\mathbb E\{\mathbf s\mathbf s^H\} = \mathbf I_{K+M}$. Then, the transmitted signal can be written as $\mathbf x = \mathbf {Ws}$, where $\mathbf W$ denotes the full digital transmit beamformer, which can be defined as,
	\begin{equation}
		\mathbf W = [\underbrace{\mathbf w_1, \dots,\mathbf  w_{K}}_\text{for communication and sensing}, \underbrace{ \mathbf w_{K+1}, \dots, \mathbf w_{K+M}}_\text{for sensing}] \in \mathbb C^{N_t\times (K+M)}.
	\end{equation}
	The received signal at user $k$ can be represented as
	\begin{equation}
		y_k = \mathbf h_k^H\mathbf w_k + \sum_{j=1,j\neq k}^{K} \mathbf h_k^H\mathbf w_j + \sum_{l=K+1}^{K+M}\mathbf h_k^H\mathbf w_l + n_k, 
	\end{equation}
	where $\mathbf h_k \in\mathbb C^{N_t\times 1}$ denotes the real channel between the BS and user $k$, and $n_k \sim \mathcal{CN}(0,\sigma_k^2)$ denotes the additive white Gaussian noise (AWGN) at user $k$. Accordingly, the received SINR at user $k$ can be derived as
	\begin{equation}
		\gamma_k = \frac{|\mathbf h_k^H\mathbf w_k|^2}{\sum_{j=1,j\neq k}^{K} |\mathbf h_k^H\mathbf w_j|^2 + \sum_{l=K+1}^{K+M}|\mathbf h_k^H\mathbf w_l|^2 + \sigma_k^2},
	\end{equation}
	with the received data rate at user $k$ obtained as
	\begin{equation}
		r_k = \log_2(1+\gamma_k).
	\end{equation}

	As for sensing, we denote $\theta_m,\,m\in\{1,\dots,M\}$ as the direction of angle (DoA) of target $m$. The communication and sensing data streams can be simultaneously employed for sensing the targets. Then, the beampattern gain at target $m$ can be written as
	\begin{equation}
		P(\theta_m) = \mathbf a^H(\theta_m)\mathbf R_{\mathbf w}\mathbf a(\theta_m), \label{Probing_pow}
	\end{equation}
	where $\mathbf a(\theta_m) = [1, e^{-j\frac{2\pi}{\lambda}d_0\sin(\theta_m)},\dots, e^{-j\frac{2\pi}{\lambda}(N_t-1)d_0\sin(\theta_m)}]$ is the steering vector towards angle $\theta_m$, $\mathbf R_{\mathbf w} = \mathbb E\{\mathbf x\mathbf x^H\} =\mathbf W\mathbf W^H$ is the covariance matrix of the transmitted signal.
	
	\subsection{CSI and Target Location Uncertainty Model}
	
	The real channel $\mathbf h_k$ can be modeled as 
	\begin{equation}
		\mathbf h_k = \hat{\mathbf h}_k + \Delta\mathbf h_k,\;\forall k, \label{chan_model}
	\end{equation}
	where $\hat{\mathbf h}_k$ denotes the estimated channel, and $\Delta\mathbf h_k$ is the CSI error following the bounded error model as \cite{XH}
	\begin{equation}
		\Vert \Delta\mathbf h_k\Vert_2 \le \epsilon_k,\;\forall k.
	\end{equation}
	
	Because we only focus on the angle of direction (AoD) of the point-like target under far-field assumption, consider the deviation of the target AoD, and the steering vector of target $m$ can be modeled as 
	\begin{equation}
		\tilde{\mathbf a}_m = \left\{ \mathbf a(\theta_m)|\theta_m \in [\theta_{m,min}, \theta_{m,max}] \right\},\; \forall m,
	\end{equation}
	where $\theta_{m,min}$ and $\theta_{m,max}$ denote the minimum and the maximum values of possible $\theta_m$, respectively.
	
	\subsection{Problem Formulation}
	
	Under the existence of CSI and AoD uncertainty in the ISAC system, we aim to propose a framework to maximize the weighted sum of worst-case user data rate and the beampattern gain, by optimizing the digital beamforming such that
 \begin{small}
	\begin{align}
		(\text{P1})\;\max_{\mathbf W}&\; \rho\left(\min_{\{\Delta\mathbf h_k\}}\sum_{k=1}^K r_k\right)   + (1-\rho)\min_{\{\tilde{\mathbf a}_m\}}\sum_{m=1}^M P(\theta_m), \label{P1obj}\\
		\text{s.t. } & \Vert\mathbf W\Vert_F^2 \le P_0, \tag{\ref{P1obj}a} \label{Cons_pow}
	\end{align}
 \end{small}%
	where $\rho$ is the weight allocated to communication function. Constraint (\ref{Cons_pow}) denotes the power budget at the BS. The main challenges to solve this problem are i) the non-smooth objective function due to max-min problem; ii) the intractable CSI and angle uncertainty; iii) the non-convex expression of $r_k$ and $P(\theta_m)$ as the objective function.
	
	\section{Robust Beamforming Optimization}
	
	In this section, we provide the proposed solution for the robust beamforming design. Since constraint (\ref{Cons_pow}) is already convex, we focus on dealing with the non-convex objective function. \red{Specifically, (P1) can be separated to communication and sensing parts as
    \begin{small}
	\begin{align}
		(\text{P1-a})\;\max_{\mathbf W}&\; \rho\left(\min_{\{\Delta\mathbf h_k\}}\sum_{k=1}^K r_k\right), \label{P1-a-obj}\\
		\text{s.t. } & (\ref{Cons_pow})),\notag
	\end{align}
    \end{small}%
    \begin{small}
	\begin{align}
		(\text{P1-b})\;\max_{\mathbf W}&\;  (1-\rho)\min_{\{\tilde{\mathbf a}_m\}}\sum_{m=1}^M P(\theta_m), \label{P1-b-obj}\\
		\text{s.t. } & (\ref{Cons_pow})).\notag
	\end{align}
    \end{small}%
    Then, we will discuss the solutions to (P1-a) and (P1-b) in section III.A and section III.B, respectively.
    }
    
	\subsection{Solving (P1-a) with S-Procedure}
	
	Focusing on communication, by introducing auxiliary variable $\delta_k$, (P1-a) can be further transformed as
    \begin{small}
	\begin{align}(\text{P1-a1})\,
		\max_{\mathbf W,\bm\delta}&\;  \rho\sum_{k=1}^K\log_2(1+\delta_k), \label{P1-a1-obj}\\
		\text{s.t. } & 
		\min_{\Delta\mathbf h_k} \gamma_k\ge \delta_k,\;\forall k,  \label{Cons_rate} \tag{\ref{P1-a1-obj}a}\\
		& (\ref{Cons_pow}), \notag
	\end{align}
    \end{small}%
	where (\ref{Cons_rate}) is equivalent to 
	\begin{gather}
		|\mathbf h_k^H\mathbf w_k|^2 \ge \nu_k, \;\forall \Vert\Delta\mathbf h_k\Vert_2 \le \epsilon_k,\,\forall k, \label{Cons_SINR1} \\
		\sum_{j=1,j\neq k}^{K+M}|\mathbf h_k^H\mathbf w_j|^2 +\sigma_k^2\le \beta_k, \;\forall \Vert\Delta\mathbf h_k\Vert_2 \le \epsilon_k,\,\forall k, \label{Cons_SINR2} \\
		\frac{\nu_k}{\beta_k} \ge \delta_k,\,\forall k, \label{Cons_SINR3}
	\end{gather}
	where $\bm\beta = [\beta_1, \dots, \beta_K]^T$ and $\bm\nu = [\nu_1,\dots,\nu_K]^T$ are slack variables. The main challenge introduced by the communication measurement is to tackle the above three non-convex constraints.
	
	Recalling (\ref{chan_model}), (\ref{Cons_SINR1}) can be rewritten as
	\begin{equation}
		\begin{split}
			\Delta\mathbf h_k^H&\mathbf w_k\mathbf w_k^H\Delta\mathbf h_k + 2\mathcal Re\{\hat{\mathbf h}_k^H\mathbf w_k\mathbf w_k^H\Delta\mathbf h_k\}  \\
			&+ \hat{\mathbf h}_k^H\mathbf w_k\mathbf w_k^H\hat{\mathbf h}_k \ge \nu_k, \;\forall \Vert\Delta\mathbf h_k\Vert_2 \le \epsilon_k,\,\forall k, \label{SINR1_temp1}
		\end{split}
	\end{equation}
	where the quadratic expression with respect to variable $\mathbf w_k$ makes the constraint non-convex. Thus, this problem can be solved iteratively by successive convex approximation (SCA) method. The lower bound of the left-hand-side can be approximated by its first Taylor expansion, and (\ref{SINR1_temp1}) can be further transformed as
	\begin{equation}
		\Delta\mathbf h_k^H\mathbf\Lambda_k\Delta\mathbf h_k + 2\mathcal Re\{\mathbf b_k^H\Delta\mathbf h_k\} + c_k \ge \nu_k, \;\forall \Vert\Delta\mathbf h_k\Vert_2 \le \epsilon_k,\,\forall k, \label{SINR1_temp2}
	\end{equation}
	where $\mathbf\Lambda_k$, $\mathbf b_k$ and $c_k$ can be obtained by
	\begin{gather}
		\mathbf \Lambda_k = \mathbf w_k\mathbf w_k^{(t)H} + \mathbf w_k^{(t)}\mathbf w_k^H - \mathbf w_k^{(t)}\mathbf w_k^{(t)H}, \\
		\mathbf b_k = \mathbf w_k\mathbf w_k^{(t)H}\hat{\mathbf h}_k + \mathbf w_k^{(t)}\mathbf w_k^H\hat{\mathbf h}_k - \mathbf w_k^{(t)}\mathbf w_k^{(t)H}\hat{\mathbf h}_k, \\
		c_k = \hat{\mathbf h}_k^H\left(\mathbf w_k\mathbf w_k^{(t)H} + \mathbf w_k^{(t)}\mathbf w_k^H - \mathbf w_k^{(t)}\mathbf w_k^{(t)H}\right)\hat{\mathbf h}_k,
	\end{gather}
	where $\mathbf w_k^{(t)}$ denotes the results obtained at the last iteration.
	
	Next, the S-Procedure is employed to deal with the bounded CSI error model, and (\ref{SINR1_temp2}) is equivalent to
	\begin{equation}
		\begin{bmatrix}
			\mathbf \Lambda_k + \varphi_k\mathbf I_{N_t}  & \mathbf b_k \\
			\mathbf b_k^H & c_k - \nu_k -\varphi_k\epsilon_k^2		
		\end{bmatrix} \succeq \mathbf 0,\,\forall k, \label{Cons_SINR1_proc}
	\end{equation}
	where $\bm\varphi = [\varphi_1, \dots, \varphi_K]$, $\varphi_k \ge 0$ are slack variables.
	
	As for (\ref{Cons_SINR2}), first we apply Schur complement to transform (\ref{Cons_SINR2}) as
	\begin{small} 
		\begin{equation}
			\begin{bmatrix}
				\beta_k-\sigma_k^2 & (\hat{\mathbf h}_k^H + \Delta \mathbf h_k^H)\mathbf W_{\bar{k}}\\ \mathbf W_{\bar{k}}^H(\hat{\mathbf h}_k + \Delta \mathbf h_k) & \mathbf I_{N_t}
			\end{bmatrix} \succeq \mathbf 0,\forall \Vert\Delta\mathbf h_k\Vert_2 \le \epsilon_k,\forall k,\label{SINR2_temp1}
		\end{equation}
	\end{small}%
	where $\mathbf W_{\bar k}$ is stacked as
	\begin{equation}
		\mathbf W_{\bar k} = [\mathbf w_1,\dots,\mathbf w_{k-1}, \mathbf w_{k+1}, \dots, \mathbf w_{K+M} ]. 
	\end{equation}
	Then, (\ref{SINR2_temp1}) can be further equivalently transformed as \cite{9110587}
	\begin{equation}
		\begin{bmatrix}
			\beta_k - \sigma_k^2 - \xi_k & \hat{\mathbf h}_k^H\mathbf W_{\bar k} & \mathbf 0_{1\times N_t} \\
			\mathbf W_{\bar k}^H\hat{\mathbf h}_k & \mathbf I_{K+M-1} & \epsilon_k\mathbf W_{\bar{k}}^H \\
			\mathbf 0_{N_t\times 1} & \epsilon_k\mathbf W_{\bar k} & \xi_k\mathbf I_{N_t}
		\end{bmatrix} \succeq \mathbf 0,\; \forall k, \label{Cons_SINR2_proc}
	\end{equation}
	where $\bm \xi = [\xi_1, \dots, \xi_K]$, $\xi_k \ge 0$ are the slack variables.
	
	To address (\ref{Cons_SINR3}), we first rewrite it as
	\begin{equation}
		\nu_k \ge \delta_k\beta_k = \frac{1}{4}\left( (\delta_k+\beta_k)^2 - (\delta_k-\beta_k)^2 \right),\,\forall k,
	\end{equation}
	where the two slack variables at the right-hand-side of the inequality are decoupled by rewriting it as a difference of convex function. We further linearly approximate its upper bound, and then (\ref{Cons_SINR3}) is obtained as
	\begin{small}
		\begin{equation}
			\nu_k \ge \frac{1}{4}\left( \left(\delta_k+\beta_k\right)^2 - 2\left(\delta_k^{(t)}-\beta_k^{(t)}\right)(\delta_k-\beta_k) + \left(\delta_k^{(t)}-\beta_k^{(t)}\right)^2  \right), \forall k. \label{Cons_slack}
		\end{equation}
	\end{small}%
	
	Based on the above derivations, problem (P1-a) can be reformulated as
	\begin{align}
		(\text{P1-a2})\,\max_{\mathbf W,\bm\delta,\bm\beta,\atop\bm\varphi,\bm\xi,\bm\nu}&\; \rho\sum_{k=1}^K\log_2(1+\delta_k)   , \label{P1-a2-obj}\\
		\text{s.t. } &  (\ref{Cons_pow}),(\ref{Cons_SINR1_proc}),(\ref{Cons_SINR2_proc}),(\ref{Cons_slack}). \notag
	\end{align}

	\subsection{Solving (P1-b) with Convex Hull}
	
	Focusing on (P1-b), to make the sensing beampattern gain more tractable, (\ref{Probing_pow}) can be rewritten as
	\begin{equation}
		P(\theta_m) = \tr\left(\mathbf A(\theta_m)\mathbf R_{\mathbf w}\right),
	\end{equation}
	where $\mathbf A(\theta_m) = \mathbf a(\theta_m)\mathbf a(\theta_m)^H$. Because we only focus on sensing in this stage, temporarily ignoring communication metric and constant $(1-\rho)$, problem (P1-b) can be reduced as
	\begin{align}
		(\text{P1-b1}) \; &\max_{\mathbf W} \min\limits_{\mathbf A(\theta_m)\in \tilde{\mathbf A}(\theta_m) } \sum\limits_{m=1}^M \tr\left(\mathbf A(\theta_m)\mathbf R_{\mathbf w}\right) \\
        &\text{s. t. } (\ref{Cons_pow}), \notag
	\end{align}
	where $\tilde{\mathbf A}(\theta_m) = \{ \mathbf a(\theta_m)\mathbf a^H(\theta_m) |\theta_m \in [\theta_{m,min}, \theta_{m,max}]\},\forall m$. To make it tractable, we construct a convex hull of $\tilde{\mathbf A}(\theta_m)$ as \cite{8334240}
	\begin{equation}
		\mathbf B_m = \left\{\sum_{s=1}^{S_k}\mu_{m,s}\mathbf A_s(\theta_m) | \sum_{s=1}^{S}\mu_{m,s} = 1, 0\le\mu_{m,s}\le 1\right\}, \;\forall m,
	\end{equation}
	where $\mu_{m,s}$ is the weight, $S_k$ is the total number of samples, and $\mathbf A_s(\theta_m)$ is the $s$-th sample. By focusing on the sum of beampattern gain in the objective function, and ignoring the constraints discussed in the last part, we have the following proposition.
	
	\emph{Proposition 1}:  To maximize the the sum of minimization over the set $\tilde{\mathbf A}(\theta_m)$ is equivalent to that over $\mathbf B_m$, which can be represented as
	\begin{equation}
		\begin{split}
			\max\limits_{\mathbf W}\min\limits_{\mathbf A(\theta_m) \in \tilde{\mathbf A}(\theta_m)} \sum\limits_{m=1}^M \tr\left(\mathbf A(\theta_m)\mathbf R_{\mathbf w}\right) \\
			=  \min\limits_{\mathbf A(\theta_m) \in \mathbf B_m}\max\limits_{\mathbf W} \sum\limits_{m=1}^M \;\tr\left(\mathbf A(\theta_m)\mathbf R_{\mathbf w}\right),
		\end{split} \label{beampattern_obj}
	\end{equation}
	
	\emph{Proof}: The proof can be referred to \cite{8334240}.
	
	Thus, problem (P1-b1) is equivalent to
	\begin{align}
		(\text{P1-b2})\;&\min_{\{\mu_{m,s}\}}\max_{\mathbf W}\sum_{m=1}^M \sum_{s=1}^{S}\mu_{m,s}\tr\left( \mathbf A_s(\theta_m)\mathbf R_{\mathbf w} \right)
		\label{beampattern_obj2} \\
        & \text{s. t. } (\ref{Cons_pow}). \notag
	\end{align}
	Then, we will optimize $\mathbf W$ and $\{\mu_{m,s}\}$ in two layers. 
	
	In the outer layer, assuming $\mathbf W$ is fixed, the reverse H$\ddot{\text o}$lder inequality can be applied as
	\begin{small}
	\begin{equation}
		\begin{split}
			\sum_{s=1}^{S}&\mu_{m,s}\tr\left( \mathbf A_s(\theta_m)\mathbf R_{\mathbf w} \right) \\
			&\ge \left(\sum_{s=1}^S(\mu_{m,s})^\frac{1}{2}\right)^2\left(\sum_{s=1}^S\left(\tr\left( \mathbf A_s(\theta_m)\mathbf R_{\mathbf w}\right)\right)^{-1}\right)^{-1},\forall m,
		\end{split}
	\end{equation}
	\end{small}%
	where the equality holds if and only if $\frac{\mu_{m,1}^\frac{1}{2}}{\left(\tr\left( \mathbf A_s(\theta_m)\mathbf R_{\mathbf w}\right)\right)^{-1}} = \dots = \frac{\mu_{m,S}^\frac{1}{2}}{\left(\tr\left( \mathbf A_S(\theta_m)\mathbf R_{\mathbf w}\right)\right)^{-1}}$. Together with $\sum_{s=1}^{S}\mu_{m,s} = 1, 0\le\mu_{m,s}\le 1$, the $\mu_{m,s}$ that results in the worst case beampattern gain can be computed as
	\begin{small}
	\begin{equation}
		\mu^{*}_{m,s} = \frac{\left(\tr\left( \mathbf A_s(\theta_m)\mathbf R_{\mathbf w}^{(t)}\right)\right)^{-2}}{\sum_{s=1}^S\left(\tr\left( \mathbf A_s(\theta_m)\mathbf R_{\mathbf w}^{(t)}\right)\right)^{-2}}, \;\forall m, \label{opt_mu}
	\end{equation}
	\end{small}%
	where $(t)$ denotes the optimal value obtained at the last iteration.

	In the inner layer, with fixed $\{\mu_{m,s}\}$, (P1-b2) can be reformulated as
	\begin{align}
		(\text{P1-b3}) \;&\max_{\mathbf W}\sum _{m=1}^M\sum_{n=1}^{K+M} \mathbf w_j^H\bar{\mathbf B}_m\mathbf w_j \label{beampattern_obj3} \\
        &\text{s. t. } (\ref{Cons_pow}),\notag
	\end{align}
	where $\bar{\mathbf B}_m = \sum_{s=1}^S\mu_{m,s}^{*(t)}\mathbf A_s(\theta_m)$. To solve this non-convex problem, SCA method can be utilized again to approximate the objective function by its first Taylor expansion as
	\begin{equation}
		\mathbf w_j^H\bar{\mathbf B}_m\mathbf w_j \ge 2\mathcal Re\left\{\mathbf w_j^{(t)H}\bar{\mathbf B}_m\mathbf w_j\right\} - \mathbf w_j^{(t)H}\bar{\mathbf B}_m\mathbf w_j^{(t)}.
	\end{equation}
    Then with the approximation, (P1-b3) is transformed to be a simple convex problem. \red{Because (P1-a2) about communication is unrelated to updating $\mu_{m,s}$, we update beamforming matrix $\mathbf W$ in the inner layer, by combining the communication part solving (P1-a2) in section III.A with sensing part solving the transformed (P1-b3) as}
	\begin{small}
	\begin{align}
		& (\text{P2})\;\max_{\mathbf W,\bm\delta,\bm\beta\atop\bm\varphi,\bm\xi,\bm\nu}\; \rho\sum_{k=1}^K\log_2(1+\delta_k)  + (1-\rho)\notag \\
		&\times\sum _{m=1}^M\sum_{j=1}^{K+M}\left(2\mathcal Re\left\{\mathbf w_j^{(t)H}\bar{\mathbf B}_m\mathbf w_j\right\} - \mathbf w_j^{(t)H}\bar{\mathbf B}_m\mathbf w_j^{(t)}\right) \label{Obj_W}\\
		& \text{s.t. }  (\ref{Cons_pow}), (\ref{Cons_SINR1_proc}),(\ref{Cons_SINR2_proc}),(\ref{Cons_slack}), \notag
	\end{align}
	\end{small}%
	which is an semidefinite programming (SDP) problem that can be efficiently solved by the existing toolbox such as CVX.
	
	
	\subsection{Overall Algorithm}
	
	Overall, problem (P2) can be solved by two layers, where  $\{\mu_{m,s}\}$ is optimized in the outer layer, and $\mathbf W$ is updated in the inner layer. The overall algorithm is summarized as \textbf{Algorithm 1} and \textbf{Algorithm 2}.
	\begin{algorithm}[t]  
		\caption{Inner layer: worst-case beamforming design.} 
		\begin{algorithmic}[1]  
			\State \textbf{Initialize} $\mathbf W^{(0)}$, $\bm\delta^{(0)}$, $\bm \beta^{(0)}$. Set iteration index $t = 0$.
			\Repeat
			\State \textbf{Update} beamforming matrix $\mathbf W$ and auxiliary variables $\bm\delta$,  $\bm\beta$, $\bm\nu$, $\bm \varphi$, $\bm\xi$ by solving problem (P2) based on fixed $\mu$.
			\State Set $t = t+1$.
			\Until{Convergence.}
		\end{algorithmic}  
	\end{algorithm}
	\begin{algorithm}[t]  
		\caption{Outer layer: overall robust ISAC beamforming optimization.}  
		\begin{algorithmic}[2]  
			\State \textbf{Initialize} $\bm \mu^{(0)}$ and $\mathbf W^{(0)}$. Set iteration index $i = 0$.
			\Repeat
			\State \textbf{Update} weight coefficients for sensing steering vector $\mu_{m,s}$ by (\ref{opt_mu}).
			\State \textbf{Update} beamforming matrix $\mathbf W$ by \textbf{Algorithm 1}.
			\State Set $i = i+1$.
			\Until{Convergence.}
		\end{algorithmic}  
	\end{algorithm}

	The computational complexity of the proposed robust algorithm mainly depends on the complexity of optimizing $\mathbf W$ in \textbf{Algorithm 1} and updating $\bm\mu$ in \textbf{Algorithm 2}. In the inner loop, solving problem (P2) contains variables $n$ of order $N_tM + (N_t+5)K$, $K+1$ quadratic constraints and $K$ linear matrix inequality constraints with sizes $N_t+1$ and $N_t + K+M$, respectively. Thus, the total computational complexity of solving problem (P2) is $\mathcal O(I_I\log_2(1/\epsilon)n(K(N_t+K+M)^3 + nK(N_t+K+M)^2 + n^2))$, where $I_I$ denotes the number of iterations of the inner loop, and $\epsilon$ denotes the accuracy of the SCA based method. In the outer loop, the main complexity depends on computing $\mu_{m,s}$ with (\ref{opt_mu}), which is $\mathcal O(SN_t^3)$. Finally, the overall computational complexity of the proposed \textbf{Algorithm 2} can be represented as $\mathcal O(I_O(I_I\log_2(1/\epsilon)n(K(N_t+K+M)^3 + nK(N_t+K+M)^2 + n^2) +SN_t^3 ))$, where $I_O$ is the number of iterations of the outer loop.

	\section{Numerical Results}
	
	In this section, we provide numerical simulations to evaluate the performance of our proposed robust beamforming algorithm for the ISAC system. We assume the BS is equipped with $N_t = 8$ transmit antennas, serving $K = 3$ users and sensing $M = 2$ targets. The distances between the BS and the users are set as between 20 and 70 m. The estimated AoDs of the users and targets are $[13^\circ, 50^\circ, 65^\circ]$ and $[121^\circ, 127^\circ]$, respectively. The bound for CSI error is defined as $\epsilon_k = \varpi\Vert\hat{\mathbf h}_k\Vert_2, \forall k$, with coefficient $\varpi \in [0,1)$. The range for target AoD uncertainty is denoted as $\Delta\theta_m = \theta_{m,max} - \theta_{m,min},\forall m$. For simplicity, we assume that the angle uncertainty for each target is the same, i.e. $\Delta\theta$.

	The distance dependent path loss is $\text{PL}(d) = 30 + 10\alpha\log(d)$ dB, where $d$ denotes the distance, and $\alpha = 3$ is the path loss exponent. The noise powers are assumed to be $\sigma_1^2  = \cdots = \sigma_K^2 = -80$ dBm. To verify the effectiveness and performance improvement of our proposed robust algorithm, we have two baselines, i) non-robust: optimizing the transmit beamforming matrix with estimated user channels and target angles by SCA-based algorithm; ii) steering vector matching (SVM): aligning each column of the beamforming matrix with the array response vectors of the corresponding users and targets. 
	\begin{figure}[t]
		\centering
		\includegraphics[width=0.7\linewidth]{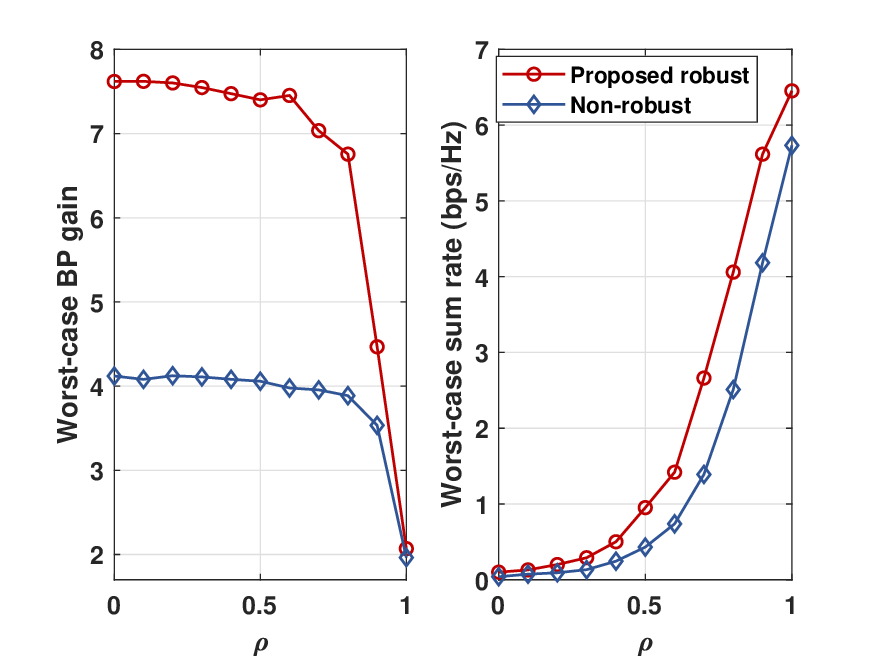}
		\caption{Worst-case sensing and communication performance versus weight $\rho$ with $P_0 = 30$ dBm.}
		\label{vsrho}
	\end{figure}
	
	\begin{figure}[t]
		\centering
		\includegraphics[width=0.7\linewidth]{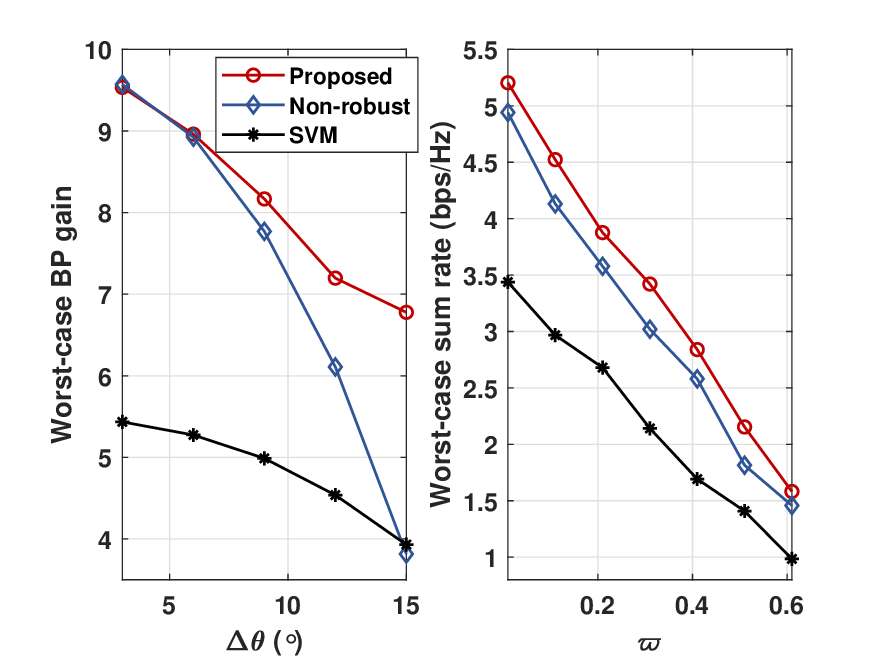}
		\caption{Worst-case sensing and communication performance versus AoD error or CSI bound with $P_0 = 30$ dBm, $\rho = 0.8$.}
		\label{vsimperfect}
	\end{figure}
	\begin{figure}[t]
		\centering
		\includegraphics[width=0.7\linewidth]{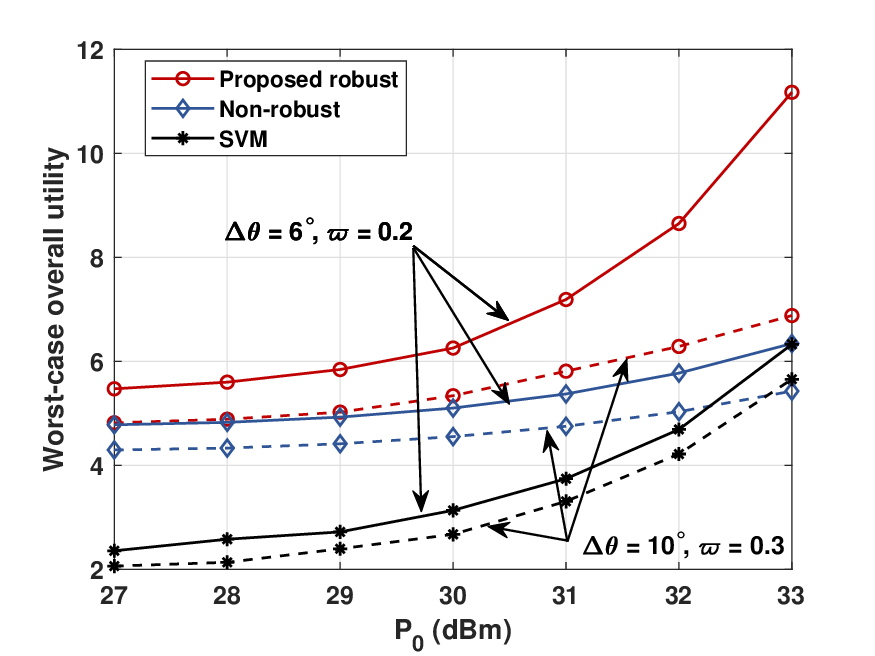}
		\caption{Worst-case dual performance versus transmit power budget with $\rho = 0.8$.}
		\label{vsP0}
	\end{figure}

	Fig. \ref{vsrho} shows the trade-off between communication and sensing performance with varying weighting factor $\rho$. We set $\Delta\theta = 15 ^\circ$ and $\varpi = 0.02$ for sensing performance analysis and $\Delta\theta = 3 ^\circ$ and $\varpi = 0.4$ for communication. It can be observed that the proposed robust algorithm significantly outperforms non-robust optimization, especially on sensing performance. This is because the proposed algorithm directly optimizes the worst-case beampattern (BP) gain in terms of sensing. Also, the trend along $\rho$ illustrates the trade-off between the two functions. The worst-case sum rate increases rapidly, while the beampattern gain decreases with the rising $\rho$,  as higher weight is assigned to communication. This feature  facilitates the design of weight allocation tailored to practical requirements.
	
	Fig. \ref{vsimperfect} shows the effect of angle and CSI uncertainty on sensing and communication performance, respectively. The proposed robust algorithm achieves higher beampattern gain and sum rate compared to the two baselines. The worst-case beampattern gain decreases with larger angle error, reflecting a similar trend in communication performance. This is intuitive that large error in AoD and channel estimation will negatively impact beamforming design. 
	
	Fig. \ref{vsP0} illustrates the worst-case overall utility (weighted sum of sensing beampattern gain and communication data rate) versus the transmit power $P_0$. The total utility increases fast  as the power budget at the base station grows. We provide two sets of angle and CSI error parameters, i.e. $\Delta\theta = 6 ^\circ,\, \varpi = 0.2$ and $\Delta\theta = 10 ^\circ,\, \varpi = 0.3$. The results indicate that the proposed robust algorithm performs better with more accurate estimation, which is consistent with the findings in Fig. \ref{vsimperfect}.

	\section{Conclusion}
	
	In conclusion, this paper introduced a dual-robust ISAC system considering imperfect CSI and uncertain target location. A two-layer iterative algorithm was proposed to address the joint sensing and communication optimization problem. Numerical results revealed that our algorithm enhanced robust performance and verified the trends in sensing and communication with varying weighting factor. Additionally, the convex hull-based worst-case sensing optimization exhibited greater potential compared to the S-Procedure-based worst-case communication optimization.

	\bibliographystyle{IEEEtran}
	\bibliography{Ref}
	
\end{document}